\documentstyle[12pt]{article}

\begin{document}

\newcommand{\NP}[1]{Nucl. \ Phys.}
\newcommand{\PL}[1]{Phys. \ Lett.}
\newcommand{\PRL}[1]{Phys.\ Rev.\ Lett. }
\newcommand{\AP}[1]{Ann.\ Phys. }
\newcommand{\IM}[1]{Inv.\ Math. }
\newcommand{\JMP}[1]{ J.\ Math.\ Phys. }
\newcommand{\RMP}[1]{Rev.\ Mod.\ Phys.}
\newcommand{\JDG}[1]{ J.\ Diff.\ Geom. }
\newcommand{\PTP}[1]{ Prog.\ Theor.\ Phys. }
\newcommand{\SPTP}[1]{Suppl.\ Prog.\ Theor.\ Phys. }
\newcommand{\PR}[1]{Phys.\ Rev. }
\newcommand{\PREP }[1]{Phys.\ Reports }
\newcommand{\NC}[1]{Nuovo \ Cim. }
\newcommand{\NCL }[1]{Nuovo\ Cim.\ Lett. }
\newcommand{\CMP}[1]{Commun.\ Math.\ Phys. }
\newcommand{\TMF}[1]{Theor.\ Math.\ Phys. }
\newcommand{\CQG}[1]{Class.\ Quant.\ Grav. }
\newcommand{\FAA}[1]{Funct. Analys. Appl. }
\newcommand{\JP}[1]{J.\ Phys. }
\newcommand{\JA }[1]{ J. Algebra }
\newcommand{\JFA}[1] {J. Funct. Anal. }
\newcommand{\MPL}[1] { Mod. Phys. Lett. }
\newcommand{\IJMP}[1] { Int. J. Mod. Phys. }
\newcommand{\RMAP}[1] { Rep.\ Math.\ Phys. }

\centerline{\Large \bf Duality in N=1, D=10 Superspace and}

\bigskip
\centerline{\Large \bf  Supergravity with Tree Level}

\bigskip
\centerline{ \Large \bf  Superstring Corrections
}

\vspace*{ 1cm}

\centerline{\bf N.A.Saulina $ ^{\it a,b}$,
M.V.Terentiev $ ^{\it a}$,  K.N.Zyablyuk $ ^{\it a}$ }

 \bigskip

\centerline{$ \it ^{ a}\ Institute \ \ Theoretical \ \
 and \ \  Experimental \ \ Physics $  \footnote{E-mail:
 saulina@vxitep.itep.ru,
 terent@vxitep.itep.ru, zyablyuk@vxitep.itep.ru
} }
\centerline{$ \it ^{ b}\  Moscow \ \ Physical \ \ Technical \ \ Institute $}

\bigskip

\begin{abstract}
The equations of motion (e.m.'s)  of the  N=1, D=10 anomaly free
supergravity, obtained in the framework of the superspace approach, are
analyzed. The formal equivalence of the usual and dual supergravities
is discussed at the level of e.m.'s. The great simplicity of the
dual formulation is established. The possibillity of the lagrangian
formulation of the dual supergravity is pointed out.
The bosonic part of the lagrangian is found.

\end{abstract}


\section{Introduction}

There are two versions of the same theory: 1) the D=10, N=1 supergravity
\cite{C}, \cite{WN}, \cite{CM} with the 3-form graviphoton field
 $H_{abc}^{(0)}$
as a member of the gravity supermultiplet, and 2) the
dual D=10, N=1 supergravity
 \cite{C}, \cite{GN1}, \cite{GN2}, \cite{BR1} where the 7-form
graviphoton field $N_{a_1\ldots a_7}$ is used instead of $H_{abc}^{(0)}$. For
further references we introduce  notations $G3$ (Gravity with the 3-form
$H$-field) and $G7$ (Gravity with the 7-form $N$-field)
 for these two versions.
The $G7$ can be derived from the $G3$-theory by
the  dual transformation at the
lagrangian  level \cite{GN1}, \cite{BR1}. Both theories are anomalous.

The connection between usual and dual versions becomes less clear if one
considers $G3$ as a low energy limit of the heterotic superstring.
 In this case superstring   corrections (anomaly cancelling) must be
added to the $H^{(0)}$ field in the $G3$-theory lagrangian \cite{GS}:
$$ H_{abc}^{(0)} \Rightarrow H_{abc} = H_{abc}^{(0)} +
 k_g\, \Omega_{abc}^{(g)}                                    \eqno(1.1)$$
where $H^{(0)} = dB $, $B$ is the two-form potential,
 $\Omega^{(g)}$ is the Lorentz-group Chern-Symons (CS) three-form, $k_g$
is a constant $\sim \alpha'$ (the string-tension constant),
 $d\Omega ^{(g)} = tr R^2 $,
where $R$ is the curvature two-form,
$trace$ is calculated in the fundamental represention of the Lorentz $O(1.9)$-
group. (We consider here the gravity sector. The incorporation of
the Yang-Mills matter can be done by standard methods).

After the change (1.1) one obtains a theory which can be made anomally-free
(by addition of special counter-terms at the one-loop level), but it is
not supersymmetric even at the tree-level.

The supersymmetric completion of such a theory has been done at the mass
shell  in papers \cite{BBLPT},
\cite{AFRR}, \cite{P} (see also \cite{MANY} for more complete list of
references). The complete lagrangian has not been constructed but it has
 become clear
that it  contains (being formulated in terms of physical fields) terms
$\sim R^2$ and an infinite number of terms $\sim k_g^q\,H^p, $
 $q\geq 1, p\geq 3 $. (Several terms of the lowest order were found in
\cite{BR1}, \cite{RW}). For brevity  we name this theory $SG3$
 (from
"Superstring inspired Gravity"). The important property of the $SG3$
is the scale invariance \cite{W}, which is the tree-level (classical)
symmetry.
 (It means that only tree-level superstring corrections are
taken into account).

We discuss  here the (scale invariant)
 dual analog of the  $SG3$ - theory. We name it
$SG7$  for short. It is  expected, that such a theory
is a low energy limit of a five-brane \cite{D}.
The $SG7$ can be formulated self-consistently and
we write explicitely the dual transformation from the $SG7$ to the
$SG3$ - theory
 at the mass-shell.
 (The inverse transformation is much more complicated and
can be defined only as a perturbative series  in $k_g$). The connection between
$SG7$ and $SG3$-theories was suggested much earlier in  \cite{AF} where
  explicit calculations were not presented
 (we agree with the
remarks from \cite{AF}).  We use the mass-shell
superspace approach to
 the problem. The iterative scheme for the dual transformation and for
the lagrangian of $SG7$ in the
component approach was discussed in \cite{BR1},\cite{BR2}.

Equations of motion (e.m.'s) in the $SG7$ are much simpler
than in the $SG3$. That makes it possible to construct a supersymmetric
lagrangian for the general $k_g\neq 0$ case. We derive here the
 bosonic part of
this lagrangian. The simplicity
of the final result is in a great contrast with the enormous complexity
of intermediate calculations. The dual
 transformation from the (relatively simple) $SG7$ to the $SG3$
 lagrangian is possible only perturbatively in $k_g$. (That explains the
complexity of the $SG3$ - theory). The fermionic and Yang-Mills matter
 sectors of
a the  $SG7$ lagrangian  can be also constructed using the
 described procedure.
 (The corresponding results will be published elsewhere).

Preparatory results for this study was given in \cite{T2},  \cite{STZ}.
Results connected with the lagrangian construction are based
on  papers \cite{Z1}, \cite{Z2}. We use the computer program "GRAMA"
\cite{ST} written in MATHEMATICA for analytical calculations in
supergravity. Our notations correspond in general to \cite{T2}
(small differences are self-evident or explained in the text).

\section{Geometrical Mass-Shell Formulation}

The superspace e.m.'s can  be formulated  universaly for the $SG3$ and the
$SG7$,
using  relations which are valid for
 both theories. These relations are:

1) Geometrical Bianchi Identities (BI's) for the supertorsion ${T_{BC}}^D$:

$$ D_{[A}{T_{BC)}}^D + {T_{[AB}}^Q\, {T_{QC)}}^D  -
{{\cal R}_{[ABC)}}^D  = 0.                 \eqno(2.1) $$

The nonzero torsion components in (2.1) are $T_{abc} \equiv \eta_{cd}
{T_{ab}}^d$ ($T_{abc}$ is a completely antisymmetric tensr),
 ${T_{ab}}^\gamma$ and:
$$ {T_{\alpha\beta}}^c = \Gamma_{\alpha\beta}^c \, , \ \ \
 {T_{a\beta}}^\gamma ={1\over 72}{({\hat T}\Gamma_a)_\beta}^\gamma
 \, , \eqno(2.2)$$
where ${\hat T} \equiv T_{abc}\Gamma^{abc} $. We use the constraints from
\cite{N}.

2) Commutation relations for supercovariant derivatives $D_A$:
$$ (D_A\, D_B - (-1)^{ab}D_B\, D_A)\, V_C =
 -\, {T_{AB}}^Q\, D_Q \,V_C -
{{\cal R}_{ABC}}^D\, V_D,                                \eqno(2.3)  $$
where $V_C$ is a vector superfield,
 ${\cal R}_{ABCD} $  is a supercurvature  (which differs in sign
 in comparison
with \cite{T2}).

3) The general result
for the spinorial derivative of the dilatino
$\chi$-superfield ($\chi_\alpha \equiv D_\alpha \phi$, where $\phi$ is
the dilaton superfield):
 $$ D_\alpha \chi_\beta = -{1\over2}\,  \Gamma^b_{\alpha \beta} D_b\phi
 +(-{1\over 36} \phi T_{abc} + A_{abc})\, \Gamma^{abc}_{\alpha \beta},
                                                            \eqno(2.4) $$
Here $A_{abc}$ is a  completely antisymmetric
superfield, which is unambiguously determined (see below) in terms
of torsion and curvature.

Some  comments on the notations are helpful. We use letters from the
beginning of the alphabet for the tangent superspace indices
$A=(a, \alpha) $  and letters from the middle of the alphabet for the
 world superspace
indices $M=(m, \mu)$. Here  $a,m$ are 10-dim.
vector indices, $\alpha, \mu$ - 16-dim. spinorial indices. The veilbein
is defined as follows \cite{BW}:
$$ {E_M}^A \vert =
\left(
\begin{array}{ll}

 {e_m}^a & \psi_m^\alpha \\
 0     & \delta_\mu^\alpha

\end{array} \right)\, ,                                   \eqno(2.5)    $$
where $\psi_m^\alpha $ ia a gravitino superfield.

The supercovariant vector derivative $D_a \equiv {E_a}^M\,D_M $ is equal to:
$$D_a = e_a^m\,D_m -\psi_a^\beta\,D_\beta \, ,              \eqno(2.6)$$
where $\psi_a = e_a^m\, \psi_m $  but
the space-time component of the covariant
derivative is:
$$ D_m \lambda = \partial\, \lambda - \omega_m \, \lambda    \eqno(2.7)  $$
where $\lambda^\gamma $ is any spinorial superfield and
 ${(\omega_m )^\beta}_\gamma \equiv {1\over 4}
{\omega_m}^{ab}{{(\Gamma_{ab})}^\beta}_\gamma   $ is the
spin-connection which is in the algebra of $O(1.9)$.

By a standard way one   finds the relation
 between the torsion-full
spin-connection in  eq.(2.7) and the standard spin-connection
$\omega_{cab}^{(0)}$ defined in terms of derivatives of $e_m^a$:
$$ \omega_{cab}= \omega_{cab}^{(0)}(e)   +{1\over2}\,T_{cab}+C_{cab}\, ,
                                                        \eqno(2.8)$$
 where:
  $$ C_{cab}= \psi_a\,\Gamma_c\,\psi_b - {3\over 2}
\psi_{[a}\,\Gamma_c\,\psi_{b]}                           \eqno(2.8)$$
 We use  the  notation $  \nabla_m   $
 for a covariant
derivative with the spin-connection $\omega_m^{(0)}$
($ \nabla_{[m}e_{n]}^a =0$).  We also define  $\nabla_a \equiv e_a^m
 \nabla_m $.  Using these notations one obtains the torsion-component
${T_{ab}}^\gamma = 2e_a^me_b^n(D_{[m}e_{n]}^\gamma) $ in the form:
$$T_{ab} = 2\,\nabla_{[a}\,\psi_{b]} - {1\over 72}(\Gamma_{[a}{\hat T} +
3\, {\hat T}\Gamma_{[a})\psi_{b]} +
{1\over2}\,(\Gamma^{cd})\,\psi_{[a}\,C_{b]cd}                \eqno(2.9)$$

 Below we use  different notations ${\cal R}_{\ldots}$ and $R_{\ldots}$ for
the curvature tensor defined in terms of  spin-connections $\omega$ and
 $\omega^{(0)}$ correspondingly ($dR=d\omega + \omega^2 $).
The complete set of e.m.'s for the  gravity supermultiplet
  derived from (2.1)-(2.4)
in \cite{T2} takes the form:

   $$  \phi L_a - D_a\chi - {1\over 36} \Gamma_a {\hat
T}\chi - {1\over 24} {\hat T}\Gamma_a\chi + {1\over 42} \Gamma_a
\Gamma^{ijk}DA_{ijk} + {1\over 7} \Gamma^{ijk}\Gamma_a DA_{ijk} = 0,
\eqno(2.10)$$

$$  D_b \Gamma^b \chi + {1\over 9}{\hat T}\chi +
 {1\over 3}\Gamma^{ijk}DA_{ijk} = 0.              \eqno(2.11)  $$

$$  D_a^2 \phi + {1\over 18}\phi (T^2) - 2\, (TA) -
  {1\over 24} D\Gamma^{ijk}DA_{ijk} = 0.             \eqno(2.12)  $$

 $$ \phi {\cal R}_{ab} -  L_{(a}\Gamma_{b)}\chi -
 {1\over 36}\phi\eta_{ab}(T^2) +
 D_{(a}D_{b)}\phi -$$
 $$-2\, (TA)_{(ab)} + {3\over 28}D{\Gamma^{ij}}_{(a}DA_{b)ij} -
 {5\over 336}\eta_{ab} D\Gamma^{ijk}DA_{ijk}=0.       \eqno(2.13)$$

$$ D_{[a}(\phi T_{bcd]}) + {3\over 2} T_{[ab}\Gamma_{cd]}\chi
 + {3\over 2}
\phi (T^2)_{[abcd]} +$$
$$+ {1\over 12} (T\epsilon A)_{abcd} + 6\,  (TA)_{[abcd]}
 + {3\over 4}D{\Gamma_{[ab}}^jDA_{cd]j} = 0.
                                                           \eqno(2.14) $$

$$ D^aT_{abc} = 0,                                        \eqno(2.15)  $$

  There are constraints:
$$T_{ab}\Gamma^{ab} =0,                                   \eqno(2.16) $$
 $$  {\cal R} - {1\over 3}\, (T^2) =0,                      \eqno(2.17)  $$
where ${\cal R}$ is the supercurvature scalar
 (${\cal R} \equiv {\cal R}_{abcd}\eta^{ac}\eta^{bd}$ )

Furthemore, there are  two equations for the $A_{abc}$-superfield. The
first one \cite{N},\cite{T2} follows from the self-consistency of
equations (2.10)-(2.15),
the second one   follows from (2.4) \cite{T2}  and  means,
 that the 1200 IR contribution to the $A$-field
spinorial derivative is equal to zero.

The following  notations were used in (2.10)-(2.18):
$$  L_a = T_{ab}\Gamma^b,  \ \ T^2 = T_{ijk}T^{ijk}, \ \  TA =
T_{ijk}A^{ijk}, \ \  (TA)_{ab} = T_{aij}{A_b}^{ij},    $$
$$   (TA)_{abcd} = T_{abj}
{A_{cd}}^j, \ \ (T\epsilon A)_{abcd} = T^{ijk}\varepsilon_{ijkabcdmns}
A^{mns}  \eqno(2.18) $$

  Spinorial derivatives of
the $A_{abc}$-
superfield can be calculated in terms of torsion and curvature.
 After that the zero superspace components
 become the e.m.'s for physical fields of the $SG3$ or $SG7$ theories.
 (We use the same notations for physical
fields and corresponding superfields expecting  that it does not lead to the
confusion).

Equations under discussion are not independent. Namely
(2.12) follows from (2.13) after contraction of $a,b$ indices, but (2.11)
follows from (2.10) after multiplication by $\Gamma^a$ matrix.

In general, neglecting Yang-Mills matter,  $A_{abc} \sim k_g $ (see below).
In the limiting case $A_{abc}=0$ these equations
 describe the pure gravity sector of the $G3$ - theory if $T_{abc} =
-(1/ \phi)\,H_{abc}$. The same equations and constraints describe the
$G7$ - theory if $T_{abc} = N_{abc} $,  where:

$$N_{abc} \equiv {1\over7!}\,{\varepsilon_{abc}}^{a_1 \ldots a_7}\,
 N_{a_1\ldots a_7}                                          \eqno(2.19) $$

Now we consider in details  the general $k_g\neq 0$- case
 starting from the $SG3$-theory.

\section{Duality on the Mass-Shell}

{\bf $SG3$  theory}

\bigskip

The $H$-superfield BI's take the form:

$$ D_{[A}H_{BCD)} +{3\over2}\,{T_{[AB}}^Q\, H_{|Q|CD)} =
  -3\, k_g \, {{\cal R}_{[AB}}^{ef}{{\cal R}_{CD)}}_{ef}    \eqno(3.1) $$
($ DH= k_g\,tr{\cal R}^2 $ in
  superform notations).
Note, that $\gamma = -2\, k_g $ in \cite{T2}.

The mass-shell solution of (3.1) which is compatible with
(2.1)-(2.4) can be obtained using the constraint
 $H_{\alpha\beta\gamma} =0$ in
the standard procedure
\cite{ADR}, \cite{RRZ}, \cite{BBLPT} . We find the nonzero components of
the $H_{ABC}$-superfield in the form:

$$ {H_{\alpha\beta}}_a=\phi\,(\Gamma_a)_{\alpha\beta}
+k_g\, U^{(g)}_{\alpha \beta a}\, ,                        \eqno(3.2a) $$
$${H_{\alpha}}_{bc}=-(\Gamma_{bc}\,\chi)_{\alpha} +
 k_g\,U^{(g)}_{\alpha bc} ,
                                                      \eqno(3.2b)  $$
$$H_{abc}=
-\phi\,T_{abc}+k_g\, U^{(g)}_{abc}                        \eqno(3.2c) $$
In this place we do not need the explicit result for the
   $U^{(g)}_{\alpha \beta a}$ and
 $U^{(g)}_{\alpha bc}$ superfields (it will be
presented elsewhere). The
  $U^{(g)}_{abc}$-superfield is equal to:

$$ U^{(g)}_{abc} = -2\, D_j^2\,T_{abc} + 4\, (T^3)_{abc}
+ {2\over 27}\, (T^2)\, T_{abc}
-6\,{T_{ab}}^j\,{\cal R}_{cj}- $$
$$-6\, {T_{a}}^{ij}\, ({\cal R}_{ij,bc} - D_i\,T_{bcj}
+ D_b\, T_{cij})
 -  T_{ij}\Gamma_{abc} T^{ij} -
12\,  T_{ja}\Gamma_b {T_{c}}^j- $$
$$ - L_j\Gamma_{abc}L^j -
12\, L_a\Gamma_b L_c  +6\,L_a T_{bc} \, , \ \ [abc]
                                      \eqno(3.3)     $$
where $[abc]$ means the antisymmetrization of the expression in
 corresponding indices, $(T^3)_{abc}= T_{aij}{T_b}^{jk}{T_{ck}}^i $.
 The $U^{(g)}_{abc}$
-superfield
was discussed earlier in \cite{BBLPT}, \cite{AFRR}, \cite{RRZ} using
another parametrization (another set of constraints) .

The $A$-superfield in (2.4)  is also determined unambiguosly
from the (2,2)-component
of the BI  (3.1) (the $(p,q)$-component of a superform contains
 $p$ bosonic and
$q$ fermionic indices):
$$ A_{abc} = k_g\,A_{abc}^{(g)}          \eqno(3.4) $$
where
$$ A^{(g)}_{abc} = -{1\over 18}\, D_j^2\,T_{abc} + {5\over 18}\, (T^3)_{abc}
+ {1\over 18\cdot 12}\, (T^2)\, T_{abc}- $$
 $$-{2\over9}\,{T_{ab}}^j\,({\cal R}_{cj} +
{5\over 8}\, (T^2)_{cj}
- {1\over 9}\, {T_{a}}^{ij}\,(-{\cal R}_{ij,bc} + {5\over 4}\, D_i\,T_{bcj}
+{5\over 4}\, D_b\, T_{cij}) -$$
$$ -{1\over 24\cdot 36}\, [(T\varepsilon T^2)_{abc} +{2\over 3}\,
(T\varepsilon D \,T)_{abc}] -
{1\over 24}\, T_{ij}\Gamma_{abc} T^{ij} +
{2 \over 9}\,  T_{aj}\Gamma_b {T_{c}}^j- $$
$$ - {7\over 8\cdot 18}\, L_j\Gamma_{abc}L^j +
{1\over 18} \,L_a\Gamma_b L_c +{4\over 9}\,L_a T_{bc} -
{1\over 9} L^j \Gamma_{ab} T_{cj}\, , \ \ [abc]
                                      \eqno(3.5)   $$
where ${X\varepsilon Y}_{abc} = X^{i_1\ldots i_k}
\varepsilon_{i_1\ldots i_k abc j_1\ldots j_p}Y^{j_1\ldots j_p}, \ \ k+p+3=10 $.

The  $A_{abc}$-superfield  defined by (3.4),(3.5) turns out to be
 a solution
of eq.'s from \cite{T2}. That provides a  good check of the result.
 \footnote{In deriving eq. (3.5)
we have corrected some
errors and misprints in \cite{T2}. Namely: 1) the
factor $(-84\cdot96)$ must be inserted into the l.h.s. of eq. (3.19) in
 \cite{T2},
 2) the coefficient
2 must be changed to 4 in  next to the last term in the r.h.s of
 eq. (3.19) in
\cite{T2},
 and 3)
the result for the $\Theta_{abcd}$-tensor (see (3.18) in \cite{T2})
 must be changed to:
$$ \theta_{abcd} = (4/3)\,D_{[a}T_{bcd]} + (64/27)\,(T^2)_{[abcd]} $$
This change is due to the fact that the term $ + {1\over 14} \,
D\Gamma^{ef}G_{e,fabcd}^{(1440)} $ was missed in the l.h.s of eq. (3.16) in
\cite{T2}. Note, that  $A^{(g)}_{abc} =-2\,L_{abc}$, where $L_{abc}$ is
determined by eq. (3.19) in \cite{T2} including  all the corrections,
 mentioned above.}

Now we are ready to discuss e.m.'s (2.10)-(2.15) in the $SG3$ - theory.
 All spinorial
derivatives can be calculated using relations from \cite{T2}.
 This work is in
progress. The analogous calculations were done in \cite{P} where another
parametrization was used. Unfortunately we are not able to use results
from \cite{P}. One needs  the expression of $T_{abc}$ in terms
 of the $H_{abc}$-
field to get the final form of equations. That  may be obtained  by
inverting of eq. (3.2c) (it can be done only perturbatively in $k_g$).
Then one gets a system of equations which is enormously complicated and
obviously untractable
\footnote{It is the reason why researh in this field, starting intensively
in 1987, was stopped during the last few years.}.

Nevertheless  one can interprete all the
 e.m.'s (2.10)-(2.15) in the $SG3$.
 Equations (2.10) -(2.13) are interpreted unamiguosly as gravitino, dilatino,
dilaton and graviton e.m.'s,
eq. (2.15) becomes  the $H$-field
e.m., but eq. (2.14)  must be the $H$-field BI. Then eq. (2.14)
 must coincide with
the (4,0)-component of the BI (3.1). That is really the case.
 Namely, substituting
(3.2c) into the (4,0)-component of (3.1) we get eq. (2.14) if the
following equation is satisfied:
$$ D_a\,U_{bcd}^{(g)} + {3\over 2} \,{T_{ab}}^e\,U_{ecd}^{(g)}
 +{3\over2}{T_{ab}}^\gamma \,
U_{\gamma cd}^{(g)}
+{1\over4} \,K_{abcd} =
 {3\over 2} \, (-2\,{{\cal R}_{ab}}^{ij}{\cal R}_{cdij})\, , \
\ [abcd]   \eqno(3.6) $$
where
$$K_{abcd} = 24 (TA^{(g)})_{abcd} +{1\over3}\,(T\varepsilon A^{(g)})_{abcd}+
3\,D{\Gamma_{ab}}^j \,DA_{cdj}^{(g)}\, , \ \ [abcd]    \eqno(3.7) $$
We have checked, calculating spinorial derivatives, that (3.6) is
satisfied identically.

Note, that (3.6)  is a (4,0)-component of a general
superform-identity \cite{BBLPT}:
$$D\,U^{(g)} + K = tr {\cal R}^2   \eqno(3.8) $$
where $U^{(g)}_{(0.3)}=K_{(0.4)}=K_{(1.3)} =0$. The (2,2), (1,3), (0,4)
-components of
(3.8)  are  satisfied because they are reduced to
that used for {\it definition} of $A$ and $U^{(g)}$- superfields.

One more remark is necessary. All the relations of the $SG3$ - theory
 are invariant under the
scale transformation \cite{W}, \cite{GN2}:
   $$ X_j \rightarrow \mu^{q_j} \, X_j                     \eqno(3.9)    $$
where $X_j$ is an arbitrary  field, but $q_j$ is a numerical
factor, which has a specific value for each field,   $\mu$ is a
 common factor.  It is a classical symmetry, because the lagrangian
 is also transformed according to (3.9) with $q=-2$.

 In the Table 1  we present  the transformation rules for
different fields
  (the numerical factors in the table are
values of $q_j$ for each field):

\centerline{Table 1}
  $$
\begin{array}{||c|c||c|c||c|c||c|c||} \hline
\phi & -1   &D_\alpha &-1/4   & T_{abc}& -{1/2} &T_{ab}^\gamma & -{3/4} \\
\hline
 e_m^a &{1/2} & A_{abc} & -3/2 &H_{abc}& -{3/2} &\psi_a^\gamma & - {1/4} \\
\hline
 D_a & -{1/2} & {{\cal R}_{ab}}^{cd} & -1 & N_{abc} & -{1/2} & \chi
   & -{5/4}       \\ \hline
  \end{array} $$

Now we come to consideration of the $SG7$-case.

\bigskip
{\bf $SG7$ theory}

\bigskip
One can interpret the same  equations (2.10)-(2.15) in terms of the
7-form graviphoton superfield $N_{A_1\ldots A_7}$.
 The BI for such a field takes
the form:

$$ D_{[A_1}N_{A_2 \ldots A_8)} + {7\over2}\, {T_{[A_1A_2}}^Q \,
N_{|Q|A_3\ldots A_8)} = 0                                  \eqno(3.10) $$
($DN = 0 $ in  superform notations).
Because of the scale invariance
(3.9) it is impossible  to  add any 8-form $\sim k_g$ constructed from
curvature into the r.h.s.
of (3.10) \cite{GN2}.

It is remarkable that the following nonzero components
 provide the solution of (3.10) which is
self-consistent with (2.1)-(2.4):

$$ N_{\alpha\beta a_1 \ldots a_5} =
 - (\Gamma_{a_1\ldots a_5})_{\alpha\beta} ,  \eqno(3.11) $$
$$ N_{abc} = T_{abc}\, ,                         \eqno(3.12)  $$
where $N_{abc}$ is defined in (2.19). This solution is valid for {\it any
$A_{abc}$-field }, in particular for that, defined by (3.4), (3.5),
 derived in the $SG3$-theory.

Using (3.12) in the equations (2.10)-(2.19)  and defining the $A_{abc}$-field
according to (3.4), (3.5) , we get the mass-shell description of the
$SG7$-theory
in a closed and relatively simple form. Eq.(2.14) becomes the
 $N_{a_1\ldots a_7}$-field e.m., but eq. (2.15) is the (8,0)-component
of the  $N$-field BI.
Using (3.12) in (3.2c)  we get the duality relation between
$H_{abc}$ and $N_{a_1\ldots a_7}$ fields.
Now we come to the discussion of the lagrangian in the $SG7$ theory.

\section{Bosonic Part of the Lagrangian}

The  lagrangian of the $SG7$-theory is equal to (we consider the
gravity sector):
$$ {\cal L}^{(g)} = {\cal L}^{(g)}_0 + k_g\, {\cal L}^{(g)}_1  \eqno(4.1) $$
where $ {\cal L}^{(g)}_0$ is the gravity part of the (anomaly full)
 lagrangian of the $G7$-theory, but   $ {\cal L}^{(g)}_1$
describes the  anomaly compensating term \cite{GS}
 and other terms,
generated by supersymmetry.

The $ {\cal L}^{(g)}_0$
has a simple form \cite{Z1}, which follows from the linearity in $\phi$ and
$\chi$ -fields of the e.m.'s (2.10)-(2.15):
$$ {\cal L}^{(g)}_0 = \phi\,({\cal R}-{1\over3}\,T^2)\,
 \vert +2\chi\,\Gamma^{ab}T_{ab}\, \vert \,
                                                            \eqno(4.2) $$
(As usual the symbol $\vert$ means  the zero superspace-component of
the superfields). The bosonic part of (4.2) takes the form:
$$ {\cal L}^{(g)}_{bos} =
 \phi\,R - {1 \over 12} \, \phi \, M_{abc}^2              \eqno(4.3) $$
where $R $ is the  curvature scalar (see the comment after eq. (2.9)), but
$$M_{abc} \equiv {1\over7!}\,
{\varepsilon_{abc}}^{a_1 \ldots a_7}\,({e_{a_1}}^{m_1}
\,\ldots {e_{a_7}}^{m_7}\, N_{m_1\ldots m_7})\, ,          \eqno(4.4) $$
where $ N_{m_1\ldots m_7}= 7\, \partial_{[m_1}\,M_{m_2 \ldots m_7]}$, and
 $M_{m_1\ldots m_6}$ is the 6-form graviphoton potential of the
$SG7$ - theory. Note, that
$$M_{abc}=
T_{abc}-{1\over2}\,\psi_f\,{{\Gamma^f}_{abc}}^d\,\psi_d  \eqno(4.5) $$
as it follows from (3.12), (2.19).

The explicit form of ${\cal L}^{(g)}_0$ with all fermionic terms
  is presented
in \cite{Z1}. (The result coincides with \cite{GN1}, \cite{BR1} after the
field
redefinition).
  The field transformation
 to the set of (primed) fields
with canonical kinetic terms has the form:
 $$ e_m^a = exp({1\over6}\phi')\, {e_m^a}', \ \ \
   \phi = exp( - {4\over3}\phi') , \ \ \ \ \chi =
  -{4\over3}exp(-{17\over12} \phi')\, \chi'  $$
 $$ \psi_m = exp({1\over 12}\phi')({\psi_m}' - {1\over6}{\Gamma_m}'\chi'),
 \ \ \  N_{abc} = -2exp(-{7\over6}\phi')\, N_{abc}'    \eqno(4.6)$$
It is the Super-Weyl transformation \cite{GV} (see \cite{STZ} for
details).

Now we come to the discussion of $k_g \,{\cal L}^{(g)}_1$-term in (4.1).
It is the property of our parametrization that
 ${\cal L}^{(g)}_1$ does not
depend of $\phi$ and $\chi$ - fields.
 It means that the scale invariance
 simplifies greatly  the possible structure of ${\cal L}^{(g)}_1$.
There are  12 possible terms:
$$ {\cal L}^{(g)}_{1, bos} = \sum_{i=1}^{12} x_i\, L_i   \eqno(4.7) $$
where $x_i$ are numbers to be determined by comparison with e.m.'s
(2.10)-(2.15), but $ L_i$ are presented in the Table 2.

\centerline{Table 2}
  $$
\begin{array}{||c|c||c|c||c|c||} \hline
i & L_i       & i  & L_i  & i & L_i \\  \hline \hline
 1 & R^2    & 5 & (M^2)\,R & 9 &  M^{abc;d}(M^2)_{abcd} \\ \hline
 2 & R_{ab}^2 & 6 &  (M^2)_{ab}\, R^{ab} & 10  & (M^2)^2 \\ \hline
3 & R_{abcd}^2 & 7 & (M^2)_{abcd}R^{abcd} & 11  &  (M^2)_{ab}^2   \\ \hline
4 &\varepsilon^{0\ldots9}R_{01bc}{R_{23}}^{bc} M_{4\ldots9} & 8
 &  M^{abc} \nabla_d\nabla^d M^{abc}     &
12 & (M^2)_{abcd}\,(M^2)^{acbd} \\ \hline
  \end{array} $$
where $(M^2) = M_{abc} M^{abc},$ \ \
 $ (M^2)_{ab} = {M_a}^{cd} M_{bcd} $ and $ (M^2)_{abcd}=
 {M_{ab}}^f M_{cdf}$.

Now we come to the determination of $x_i$ in (4.7).
All the terms, containing the $M_{abc}$- field (4.4) in the lagrangian (4.7)
  can be
easily reconstructed with the help of the simple procedure \cite{Z2}.
As was discussed before, equation (2.14) (which is the $N$-field e.m.)
is equivalent to the (4,0)-component of the
$H$-field BI.
Omitting spinorial terms,  introducing the standard
 covariant derivative $\nabla_a$ and  the curvature-tensor
 $R_{abcd}$ one can  rewrite (4,0)-component of eq. (3.1) in the form:
$$ ( H_{abc} + 3\,k_g\,(2\,T_{ija}{R_{bc}}^{ij} -
 {T_a}^{ij} T_{bij;c}
+{1\over 3}\, (T^3)_{abc}))_{;d}
 = 3\, k_g\, {R_{ab}}^{ij} R_{cdij}, \ \ [abcd]       \eqno(4.8)  $$
Then with the help of eq.'s (3.2c), (3.3) and (4.5) one can  write
 everything in terms of
the $M_{abc}$ - field. After that the terms in the lagrangian,
 containing the $M_{abc}$ - field,
are  reproduced  immediately from the l.h.s. of (4.8) which has the
desired form of a complete derivative. The term $\sim MR^2$
is reproduced from the r.h.s. of (4.8). One can not distinguish between
$ R $ and $(1/12)M^2$ on the mass shell. For this reason we are able to
 determine
by this way only $x_j, \ j=4.6,7,8,9,11,12 $ and find one relation between
$ x_j, \ j=5,10$.

The terms in (4.7), containg the $M$-field,
   were also  derived by another procedure,
 which makes it  possible to
obtain also  terms $\sim R^2$.
  Calculating the variation
of ${\cal L}^{(g)}$ over the graviton field one  must get the e.m.
 (2.13). Then,
 contracting   indices, one must get
 the dilaton e.m.  Comparing with (2.12) the result of such a variation,
 (spinorial derivatives
were explicitely calculated in (2.12)), we find
 the values of $ x_i, \ i\neq 1,5,10 $ in (4.7) and find the  relation
between $x_i, \ i=1,5,10$.
There is the complete correspondence between this calculation and
the previous one, based on eq. (4.8).

The values of $x_j$ obtained by the described procedure are presented in
Table 3.

\centerline{Table 3}
  $$
\begin{array}{||c|c||c|c||c|c||} \hline
x_1 & undetermined   &x_5 &- 2/27 - 2\,x_1/12  &  x_9  & 1/2  \\ \hline
 x_2 & 2  & x_6 & -1/2 &   x_{10} & 1/162 + x_1/144  \\ \hline
x_3 & -1  &  x_7 & 0  &x_{11} & 0     \\ \hline
x_4 & (2\cdot 6!)^{-1}  &   x_8 & -1/6  & x_{12} & -1/24 \\ \hline
  \end{array} $$

Terms  containing $x_1$ in (4.7) appear in the com\-bina\-tion
 which is the square of the constraint
(2.16).
That is the reason why
$x_1$ is undetermined by comparison with e.m.'s.

To simplify the result one can make the following
 redefinition of the dilaton field in (4.2):
$$ \phi  = \tilde \phi - k_g\,x_1\,( {\cal R}
- {1\over 3}T^2) +  k_g\, {2\over 27} (T^2)
                                                          \eqno(4.9)  $$
The second term in the r.h.s of (4.9)  leads to the cancellation of
  terms
$\sim x_1$ in (4.1). Such a redefinition does not change anything at the
mass-shell due to the constraint (2.16) (note, that neglecting fermions:
$ {\cal R} -(1/ 3)T^2 =    R -(1/ 12) M^2 $).
So one can put $x_1=0$ from the very beginning in the Table 3.

The third term in (4.9) leads to the cancellation of terms $\sim RM^2$ and
$\sim M^4$ in (4.1), so one can put $x_5 = x_{10} =0 $ in the Table 3,  using
 $\tilde \phi$ instead of $\phi$.  The third  term in the r.h.s. of (4.9)
 leads to the obvious change in the basic equation (2.4) and to the
controlable
changes in  other relations, discussed before.

    Finally,  considering
 $\tilde \phi$ as an independent variable,
one can  write the  bosonic part of the lagrangian (4.1) in the form:

$$ {\cal L}^{(g)}_{bos} = {\tilde \phi}\, (R- {1\over 12}M^2) +$$
$$+k_g\, [ 2\,R^2_{ab} -R_{abcd}^2 +
{1\over 2\cdot 6!}
\varepsilon^{abcdf_1\ldots f_6}\,R^2_{abcd}M_{f_1\ldots f_6}
-{1\over 2}\,R^{ab}(M^2)_{ab}-$$
 $$ -{1\over 6}\, M^{abc}\nabla_f\nabla^fM_{abc} +
 {1\over 2}\, M^{abc;d}(M^2)_{abcd} + {1\over 162}\, (M^2)^2
  -{1\over 24}\, (M^2)_{abcd}(M^2)_{acbd}]
                                                            \eqno(4.10) $$

Terms $\sim k_g R^2 $ and $\sim k_g M^2 $ in (4.10)
are not free from  ghosts. It
is  a  consequence of a supersymmetry because
the part of $ {\cal L}^{(g)} $  quadratic in the gravitino field
   contains ghost-full terms of the type
$k_g \psi_a\Gamma^{abc} (\nabla_d)^2 \psi_{c;b} $. (We  have not
discussed them in the present   paper for short).
  It is the ignoring of these terms
 in \cite{RW}, \cite{T4} has led  to  prediction
of the ghost-free term $(R^2_{abcd} -4R^2_{ab} + R^2)$ in the lagrangian.

The lagrangian (4.10) corresponds to the SG7-theory, which must be
supersymmetric by construction after including of fermions. It contains
anomalies, but anomaly compensating counter-terms appear only at the 8-th
order in derivatives. All such terms in the supersymmetric lagrangian
can be reconstructed iteratively in $\beta$ if one adds the term $\beta X_8$
to the r.h.s. of the BI (3.10) \cite{SS},\cite{GN2},
 where $X_8 = tr{\cal R}^4 +(1/4)(tr{\cal R}^2)^2$. In the limiting case
  $\beta=0$ the SG7 is the dual analog of SG3-theory, which is also anomaly
  full, inspite of the Green-Schwarz term in the r.h.s. of the BI (3.1).
   Anomaly compensating counter-terms in the SG3-theory
    appear at the same (8-th) order in derivatives and has never been
     supersymmetrized.

 \bigskip
Research is supported by the
ISF Grant MOY000 and by RFFI Grant 95-02-05822a

\end{document}